\documentclass[%
reprint,
superscriptaddress,
showpacs,preprintnumbers,
 amsmath,amssymb,
 aps,
]{revtex4-1}
\usepackage{color}
\usepackage{lineno}
\usepackage{graphicx}
\usepackage{dcolumn}
\usepackage{bm}
\usepackage{braket}
\newcommand{\beginsupplement}{%
        \setcounter{table}{0}
        \renewcommand{\thetable}{S\arabic{table}}%
        \setcounter{figure}{0}
        \renewcommand{\thefigure}{S\arabic{figure}}%
     }

\begin{document}

\title{Table-top NMR system for high-pressure studies with \textit{in-situ} laser heating}

\author{Thomas Meier}
   \email{thomas.meier@uni-bayreuth.de}
   \affiliation{Bavarian Geoinstitute, University of Bayreuth, D-95447 Bayreuth, Germany}

\author{Anand Prashant Dwivedi}
\affiliation{Bavarian Geoinstitute, University of Bayreuth, D-95447 Bayreuth, Germany}
  \affiliation{University of Wisconsin-Milwaukee, Milwaukee, WI 53211, USA}
  
\author{Saiana Khandarkhaeva}
  \affiliation{Bavarian Geoinstitute, University of Bayreuth, D-95447 Bayreuth, Germany}

\author{Timofey Fedotenko}
  \affiliation{Material Physics and Technology at Extreme Conditions, Laboratory of Crystallography, University of Bayreuth, D-95447 Bayreuth, Germany}                         
  
\author{Natalia Dubrovinskaia}
  \affiliation{Material Physics and Technology at Extreme Conditions, Laboratory of Crystallography, University of Bayreuth, D-95447 Bayreuth, Germany}                         

\author{Leonid Dubrovinsky}
  \affiliation{Bavarian Geoinstitute, University of Bayreuth, D-95447 Bayreuth, Germany}  

\date{\today}

\begin{abstract}
High pressure Nuclear Magnetic Resonance (NMR) is known to uncover behavior of matter at extreme conditions. However, significant maintenance demands, space requirements and high costs of superconducting magnets render its application unfeasible for regular modern high pressure laboratories. Here, we present a table-top NMR system based on permanent Halbach magnet arrays with dimensions of 25 cm diameter and 4 cm height. At the highest field of 1013 mT, $^1$H-NMR spectra of Ice VII have been recorded at 25 GPa and ambient temperature. The table-top NMR system can be used together with double sided laser heating set-ups. Feasibility of high-pressure high-temperature NMR was demonstrated by collecting $^1$H-NMR spectra of H$_2$O at 25 GPa and 1063(50) K. We found that the change in signal intensity in laser-heated NMR diamond anvil cell yields a convenient way for temperature measurements.
\end{abstract}

\pacs{}
\maketitle

\section{Introduction}

Recent developments of Diamond Anvil Cell (DAC) based NMR using two-dimensional magnetic flux tailoring Lenz lens resonator structures \cite{Meier2017} pushed the field of high pressure NMR into the multi-megabar regime (1 Mbar = 100 GPa) \cite{Meier2018}.\\
At such extreme conditions, the atomic and electronic structures of solids are significantly altered, leading to a plethora of novel physical phenomena such as the hydrogen bond symmetrisation in Ice VII governed by quantum-mechanical tunneling of protons within the H-bond network at 75 GPa \cite{Meier2018a} or the intriguing observations of the formation of a free electron hydrogen sublattice in metal hydrides above 100 GPa \cite{Meier2019a}. Furthermore, recent application of spin decoupling experiments in dense  molecular hydrogen demonstrated the possibility to utilize NMR as a structural probe yielding high resolution proton spectra in the order of a few ppm. \cite{Meier2019}.\\
Most of high pressure NMR experiments so far have been conducted at temperatures close to ambient or under cryogenic cooling\cite{Meier2017b}. An application of this method at elevated temperatures above 1000 K was considered hardly possible. This dogma is rooted in two basic ideas: 1) The signal to noise ratio in NMR scales with  $\propto T^{-3/2}$ \cite{Meier2016} ($\propto T^{-1/2}$ from thermal Johnson-Nyquist noise and $\propto T^{-1}$ from the nuclear polarization of NMR active nuclei subject to an external magnetic field $B_0$), thus leading to quickly decreasing signal intensities at high $T$, and 2) spatial requirements for NMR equipment, i.e. superconducting magnets and state of the art solid-state NMR spectrometers, effectively prohibit a combination with external electrical heating or with sensitive optical equipment necessary for in-situ laser heating in DACs.
Whereas the first point restricts an application of NMR at high temperatures in DACs to mostly hydrogen nuclei, the latter is merely rooted in technical considerations.  
\\
\begin{figure*}
  \includegraphics[width=2\columnwidth]{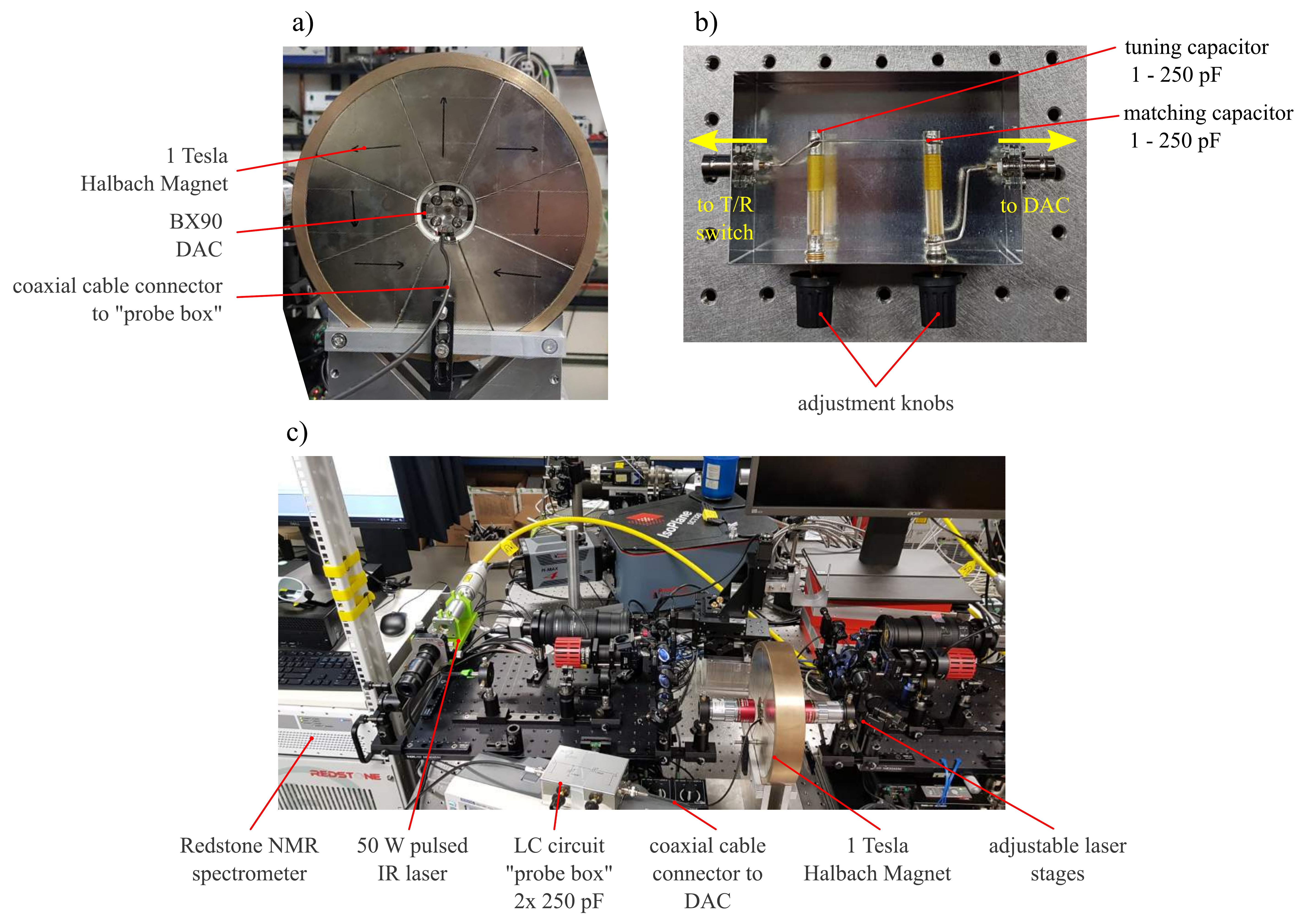}%
 \caption{Table top high pressure NMR set-up. \textbf{a)} 1 Tesla Halbach array magnet with BX90 diamond anvil cell placed in its magnetic center  \textbf{b)} photograph of the NMR probe box. \textbf{c)} laser heating set-up for \textit{in-situ} NMR measurements. 
 \label{FIG1}}
 \end{figure*} 
Over the course of the last years, novel magnet designs led to the development of table-top NMR set-ups\cite{Luy2011,Singh2016, Rachineni2017, Blumich2018}. In particular, Halbach array magnets have been established as a cost effective alternative to "standard" modern NMR magnets\cite{Moresi2003, Hills2005, Chang2006}. Halbach arrays use an array of permanently magnetised NdFeB magnets in a way that a strong and homgeneuous $B_0$ field forms in the magnets mid point. \\
Here, we present the results of the development of a low cost, space-saving table-top NMR system usable for diamond anvil cells. Furthermore, we show that this set-up allows for a combination of high pressure and high temperature \textit{in-situ} NMR experiments. 

\section{Experimental}
\subsection{DAC preparations}

Two pressure cells of type BX90\cite{Kantor2012} have been prepared with 250 $\mu$m (cell \#1) and 100 $\mu$m (cell \#2) culeted diamonds. Double stage (cell \#1) and triple stage (cell \#2) Lenz lens resonators have been prepared by initially covering each diamond with 1 $\mu$m of copper using physical vapor deposition and cutting of lens structures by means of focused ion beam milling. For electrical insulation, the rhenium gaskets have been covered by 500 nm of $Al_2O_3$ using chemical vapor deposition.\\
Excitation coils have been prepared from 80 $\mu$m thick  PTFE insulated copper wire, consisting of 4 turns and 3 mm coil diameter. These coils have been fixed on the respective diamonds` backing plate with the anvils in the center of the coils. \\
Both cells were loaded with distilled water and fine gold powder (with 1 $\mu$m particles) serving as absorber targets during NIR laser heating. Both cells have been dried to avoid suprious $^1$H-NMR signals from outside the sample chamber within the DAC. Pressure was calibrated using the shift of the first order Raman spectra of the diamond edge at the center of the anvil`s culets. Both cells were pressurized to about 25 GPa.    \\

\subsection{NMR measurements}
All NMR experiments were conducted using a specialized Halbach array magnet of about 1 T field. The magnet has a diameter of 250 mm, heigth of 40 mm and a central hole of roughly 45 mm in diameter. Similar to previous experiments on Ice VII \citep{Meier2018}, spin echo excitation was used for signal acquisition. Using nutation experiments, $\pi/2$ pulses of 3 $\mu$s for cell $\#1$ and 1 $\mu$s for cell $\#2$ at average pulse powers of about 20 W were found, in excellent agreement with previous experiments. No spurious hydrogen signals were detected. In both cases, spin lattice relaxation times were about 70 ms.  Relaxation delays were chosen to allow for full relaxation of the excited spin systems. \\
To calibrate the magnetic field along the z-axis of the Halbach array magnet (perpendicular to the magnets main plane), the position of cell $\#1$ was carefully changed in steps of 0.5 mm and the resulting center of gravity in frequency space after Fourier transform of the free induction decays were used as an internal primer for the polarizing external magnetic field $B_0$ in the sample chamber usign the relationship $\omega_0=\gamma_n B_0$. \\

\subsection{Laser Heating}
For the high temperature experiments, the Halbach magnet together with cell $\#2$ positioned at the magnetic sweet spot was placed between two objectives of a pulsed laser heating set-up described in Ref.\cite{Fedotenko2019}.  Laser heating operations were performed in continuous wave (CW) mode with nominal laser power of about 25 W. Preparation of anvils with FIB for NMR measurements significantly reduces quality of the anvil`s surface in the pressure chamber and makes the anvils almost opaque. As a result, spectroradiometic temperature measurements with such anvils became possible only at relatively high temperatures (above ~1500 K).  At temperatures below 1200 K (as in our experiments), the loss in signal-to-noise ratio (SNR) in the time domain of the free induction decay has been used (see below).

\section{Halbach array table-top high-pressure NMR system}
Figure \ref{FIG1}a shows the Halbach array permanent magnet with a BX90 pressure cell mounted in its center. A 50 $\Omega$ coaxial cable connects the excitation coil with the NMR probe box, figure \ref{FIG1}b. The probe box consists of two tuneable high power capacitors having a 250 pF sweep range for frequency tuning and impedance matching. The box is directly connected to a Redstone solid-state NMR spectrometer from \textit{Tecmag Inc}.\\
\begin{figure}[htb]
  \includegraphics[width=1\columnwidth]{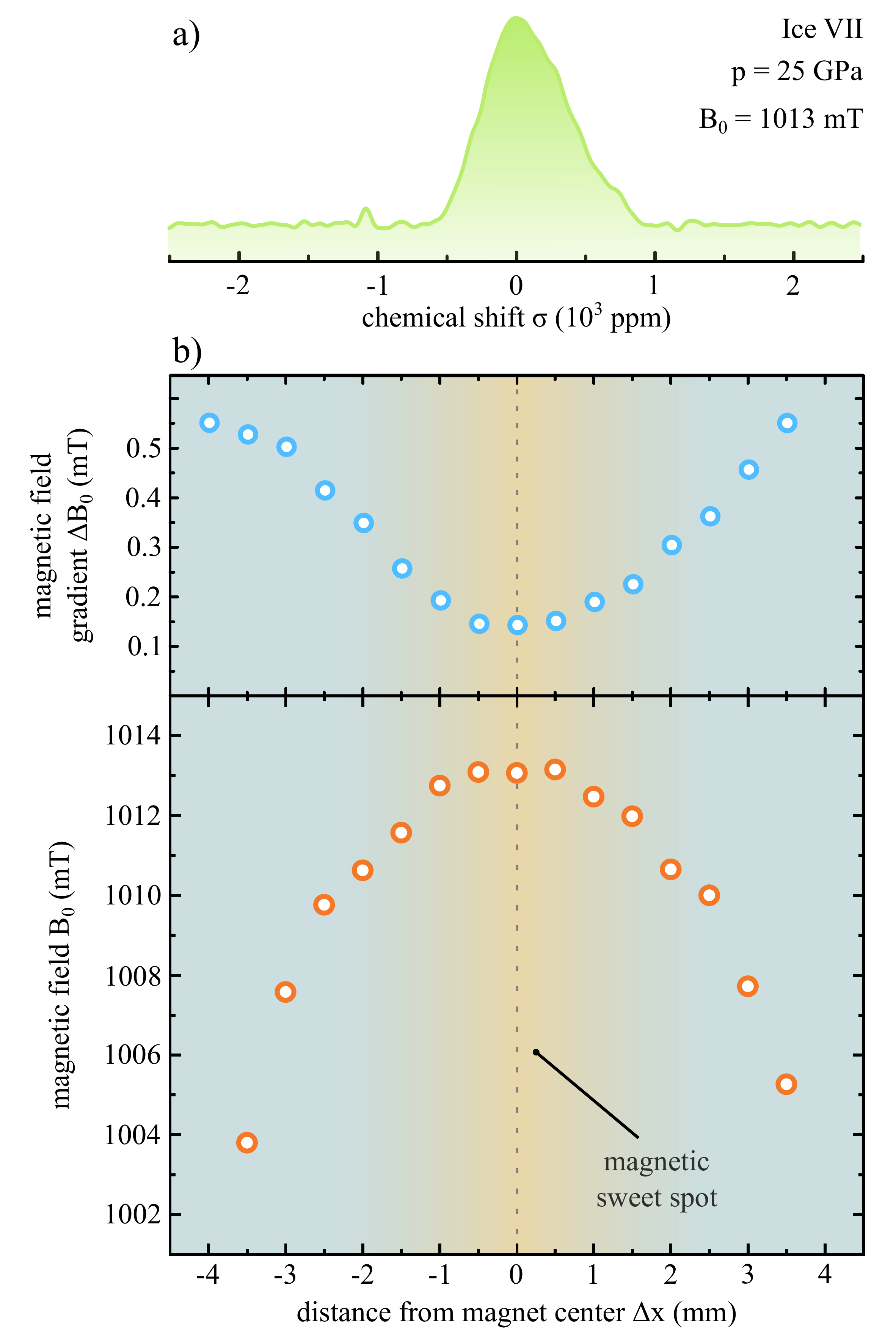}%
 \caption{a) $^1$H-NMR spectrum of Ice VII at 25 GPa and room temperature recorded in the magnetic sweet spot of the Halbach array magnet. b) magnetic field and  respective gradient in the sample cavity of the DAC at varying positions along the magnetic main axis. 
 \label{FIG2}}
 \end{figure}  
Figure \ref{FIG2} shows a $^1$H-NMR spectrum of Ice VII at 25 GPa at the highest field in the geometric center of the Halbach magnet. We found the signals' center of gravity at 43.1274 MHz, corresponding to a $B_0$ field of 1013 mT. The observed line width of about 43 kHz is similar to previous proton NMR resonances of Ice VII at this magnetic field\cite{Meier2018}.\\
Numerical simulations, supplementary figure \ref{FIGS1}, of the magnetic field of the Halbach array shows a fairly homogeneous $B_0$ field distribution in the x-y plane within a region of  $\pm$ 20 mm around its geometric center. However, as diameter and length of the inner bore of the Halbach array are comparable in size, the $B_0$ field distribution can be expected to be very inhomogeneous along the z-direction. In fact, first test measurements using a roughly 3 mm long piece of rubber in a standard RF coil yielded $^1$H-NMR line widths in the order of 20 kHz, which represents a 40 fold increase compared to a standard electromagnet. Nonetheless, given the microscopic dimensions of the sample chamber in a DAC, i.e. about 50 $\mu$m in diameter and 10 $\mu$m in thickness, these  inhomogeneity might be negligible. \\ 
Figure \ref{FIG2}b shows the magnetic field values as computed from the center of gravitites from the NMR spectra of Ice VII at 25 GPa as well as the resulting FWHM linewidths which were taken as a primer for the magnetic field gradients "felt" by the sample at the respective positions. As can be seen, even within a relatively small region of $\pm$ 4 mm from the magnet center, the magnetic field differs by 10 mT which corresponds to 426 kHz in $^1$H resonance frequencies. Moreover, at positions $\Delta x > \pm 1$mm, a significant increase in FWHM line widths was observed, using $\Delta\omega=\gamma_n \Delta B_0$ we find magnetic field gradients between 0.1 mT at $\Delta x = 0$ mm  and 0.5 mT at $\Delta x = \pm 4$mm.\\
Fortunately, a small magnetic "sweet spot" of about 1 mm near the geometric center could be found which exhibits average $B_0$ field values of 1013 mT and minimal magnetic field gradients over the sample cavity. Therefore, we can assume the magnetic field to be reasonably homogeneous at $\Delta x = 0$ for samples smaller than 40 pl which is the standard sample size for diamond anvil cell experiments in the mega-bar range. 

\section{\textit{In-situ} laser heating in NMR-DACs}

 \begin{figure}[htb]
  \includegraphics[width=0.9\columnwidth]{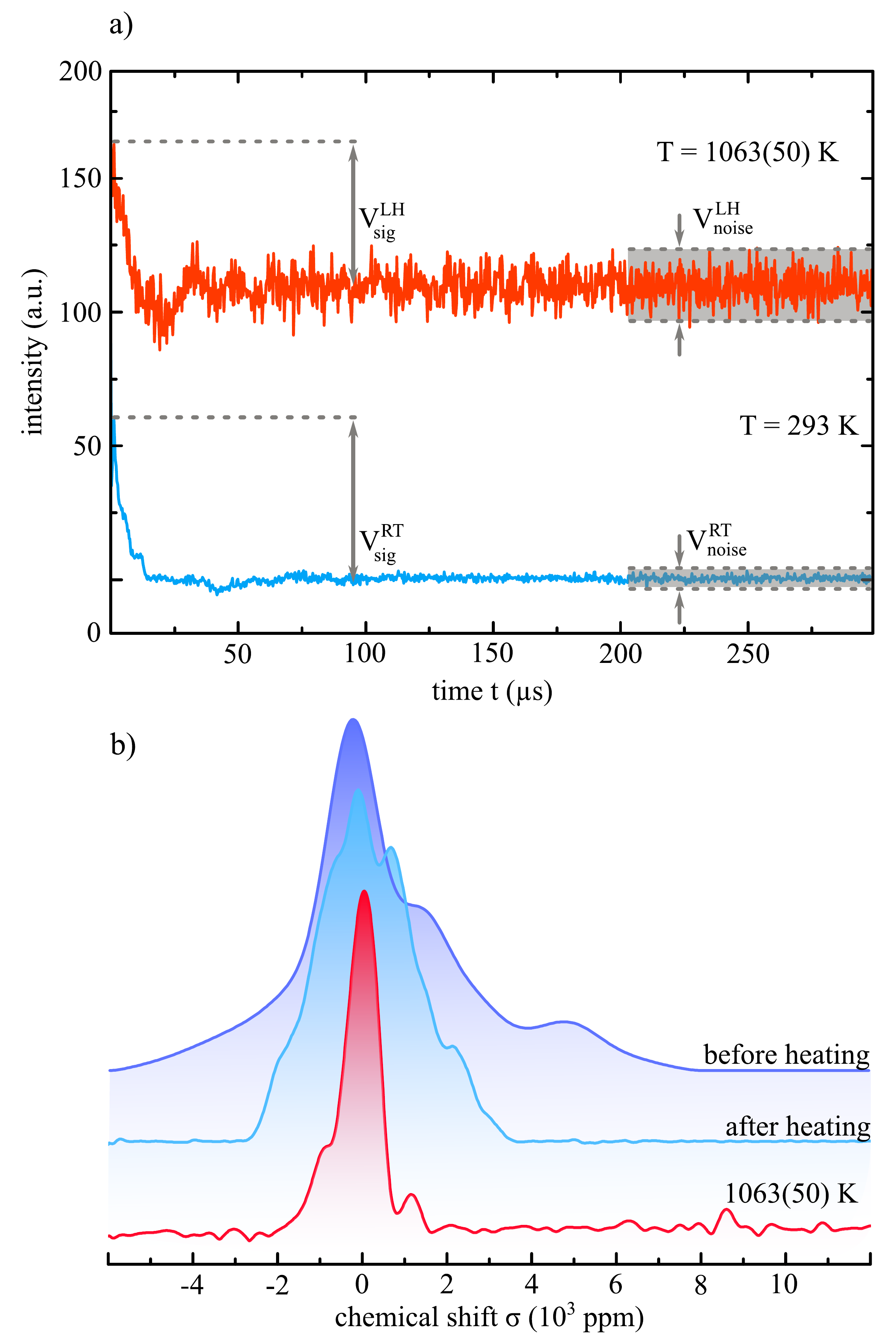}%
 \caption{\textbf{a)} Comparison of signal, $V_{sig}$, and noise voltages, $V_{noise}$ at ambient  and high temperature under continuous laser heating conditions.
\textbf{b)} $^1$H-NMR spectra of H$_2$O at 25 GPa at ambient temperature and 1063 K. 
 \label{FIG3}}
 \end{figure} 
 
Figure \ref{FIG1}c shows the table-top Halbach array set-up placed on an optical table used for double sided laser heating experiments\cite{Fedotenko2019}. First experiments have been conducted using a defocused laser spot of roughly 30 $\mu$m diameter, therefore we can assume heating of the whole sample cavity. \\
Figure \ref{FIG3}a shows a comparison between the time domain $^1$H-NMR signals at ambient and high temperatures. At room temperature, we found a signal to noise ratio (SNR) of 0.534(5) per scan. Under continuous laser heating, the SNR decreased to 0.081(9) per scan. Using the scaling relationship \cite{Fukushima1994, Mispelter2015} $SNR \propto T^{-3/2}$, the average temperature under laser heating $T_{LH}$ within the sample cavity can be estimated to be $T_{LH} = 300 K \cdot \left(SNR_{300}/SNR_{LH}\right)^{2/3} = 1063(50) K$.   \\
 Figure \ref{FIG3}b shows a comparison of the obtained $^1$H-NMR spectra at ambient temperatures and at elevated temperatures.  At $1063(50)$K a 100 ppm broad hydrogen resonance was observed at a chemical shift of $\sigma$ = 80 ppm.  \\
 After heating, the hydrogen resonances at room temperature were reinvestigated and found to be narrowed by approximately a factor of two. This effect likely originates from thermal relaxation of stress distributions present before heating and caused by pressure gradients across the Ice VII sample at 25 GPa.   \\
 
\section{Conclusions}

The aim of this work was the development of a low cost table-top NMR system which can be coupled with other high pressure instrumentation, in particular with laser-heating set-ups.\\
Combining the introduced Halbach array magnet with a small or medium sized NMR spectrometer, significantly reduces space requirements of the whole NMR set-up. Furthermore, the permanent Halbach magnet is does not require maintenance in contrast to high electrical costs of electromagnets- or regular refill of cryogenic liquids in super-conducting magnets. The introduced set-up was shown to yield $^1$H-NMR signals of similar spectral resolution in DAC-based experiments as standard electro-magnets operating at the same magnetic fields of around 1 Tesla.\\
First application within a double sided laser heating system demonstrated the potential of these set-ups for \textit{in-situ} simultaneous high pressure and high temperature NMR studies.\\
While the found spectral resolutions are too low for structural determinations, the potential scope of this new method is immense. Possible experiments include the investigation of the electronic properties of super-ionic ices which could have a significant impact in the field of geo-and planetary sciences, or the observation of formation of reaction products during laser heating assisted sample synthesis. Another field of application would be comprise the examination of melting and freezing dynamics, which would influence NMR spectra on the most basic level.\\
This novel development might lead to a wider applicability of high pressure NMR in  high pressure solid state physics, chemistry and geosciences communities.
 
\section{Acknowledgements}
The authors thank the German Research Foundation (Deutsche Forschungsgemeinschaft, DFG, projects DU 954/11-1, DU 393/13-1, DU 393/9-2, and ME 5206/3-1) and the Federal Ministry of Education and Research, Germany (BMBF, grant no. 05K19WC1) for financial support. 


\section{Supplementary Material}

\beginsupplement

\begin{figure}[htb]
  \includegraphics[width=1\columnwidth]{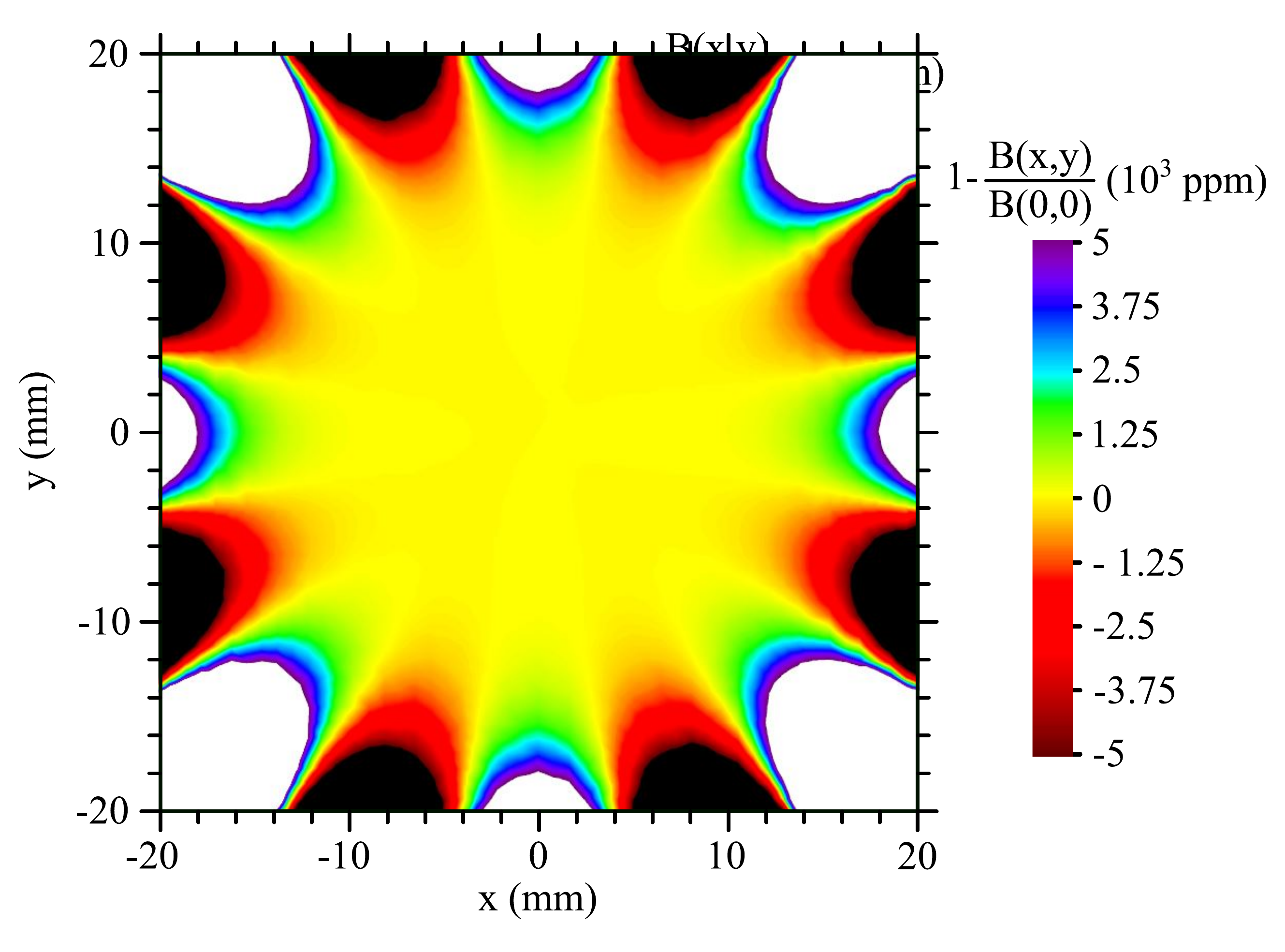}%
 \caption{Finite element simulation of the RF B$_1$ field generated by the 3rd stage Lenz lenses across a 12 pl sample cavity. The green lines denote an average value of about $\braket{B_1} = 17~ mT$. The red lines denote the upper and lower (defined by the standard deviation) boundaries of the effective volume $V_{eff}$.  
 \label{FIGS1}}
 \end{figure}



\end{document}